\documentclass[referee, sn-mathphys]{sn-jnl}
\usepackage{graphicx}
\usepackage{amsmath}
\usepackage{amssymb}

\jyear{2021}%

\begin{document}

\title[New charged anisotropic solution on paraboloidal spacetime]{New charged anisotropic solution on paraboloidal spacetime}


\author[1]{\fnm{Rinkal} \sur{Patel}}\email{rinkalpatel22@gmail.com}
\equalcont{These authors contributed equally to this work.}

\author[2]{\fnm{B. S.} \sur{Ratanpal}}\email{bharatratanpal@gmail.com}
\equalcont{These authors contributed equally to this work.}

\author*[3]{\fnm{D. M.} \sur{Pandya}}\email{dishantpandya777@gmail.com}
\equalcont{These authors contributed equally to this work.}

\affil[1]{\orgdiv{Department of Applied Science \& Humanities}, \orgname{Parul University}, \orgaddress{\street{Limda}, \city{Vadodara}, \postcode{391 760}, \state{Gujarat}, \country{India}}}

\affil[2]{\orgdiv{Department of Applied Mathematics}, \orgname{The Maharaja Sayajirao University of Baroda}, \orgaddress{\street{Faculty of Technology \& Engineering}, \city{Vadodara}, \postcode{390 001}, \state{Gujarat}, \country{India}}}

\affil*[3]{\orgdiv{Department of Mathematics, School of Technology}, \orgname{Pandit Deendayal Energy University}, \orgaddress{\street{Raisan}, \city{Gandhinagar}, \postcode{382426}, \state{Gujarat}, \country{India}}}

\abstract{\noindent New exact solutions of Einstein's field equations for charged stellar models by assuming linear equation of state  $ P_r=A(\rho-\rho_{a}) $, where $ P_r $ is the radial pressure and $ \rho_{a} $ is the surface density. By assuming $ e^{\lambda}  = 1+\frac{r^2}{R^2} $   for metric potential. The physical acceptability conditions of the model are investigated, and the model is compatible with several compact star candidates like 4U 1820-30, PSR J1903+327, EXO 1785-248, Vela X-1, PSR J1614-2230, Cen X-3. A noteworthy feature of the model is that it satisfies all the conditions needed for a physically acceptable model.}




\maketitle

	\section{Introduction}
\label{sec:1}
\noindent Einstein's field equations are a system of highly non-linear partial differential second-order equations. They are essential for modeling relativistic compact objects such as dark energy stars, gravastars, quark stars, black holes and neutron stars. The pressure distribution in the star may not be isotropic when the matter distributions have a high density in the nuclear regime, has been presented by Ruderman \cite{ruderman1972pulsars}  and Canuto \cite{canuto1974equation}. Bowers and Liang \cite{bowers1974anisotropic}  discussed the different causes for anisotropy. Since then, many researchers have studied the anisotropy Maharaj and Maartens \cite{maharaj1989anisotropic},  Gokhroo and Mehra \cite{gokhroo1994anisotropic}, Patel and Mehta \cite{patel1995exact}, Tikekar and Thomas \cite{tikekar1998relativistic}, Tikekar and Thomas \cite{tikekar1999anisotropic}, Tikekar and Thomas \cite{tikekar2005relativistic}, Thomas and Ratanpal \cite{thomas2007non}, Dev and Gleiser \cite{dev2002anisotropic}, Dev and Gleiser \cite{dev2003anisotropic}, Dev and Gleiser \cite{gleiser2004anistropic} to name a few. A large number of researchers worked on Einstein's field equations, making different assumptions in the physical content as well as spacetime metric viz., Sharma and Ratanpal \cite{sharma2013relativistic}, Murad and Fatema \cite{murad2013family}, Murad and Fatema \cite{murad2014some}, Murad and Fatema \cite{murad2015some},  Pandya {\it et al.} \cite{pandya2015modified}, Pandya and Thomas \cite{thomas2015compact}, Pandya and Thomas \cite{thomas2015new}, Ratanpal {\it et al.} \cite{ratanpal2015new}.

For constructing relativistic compact star models, an equation of state for the matter content is substituted in Einstein's field equations. Researchers used generic barotropic equations of state, in which pressure and density have linear, quadratic or polytropic relationships in several modern literary works. Sharma and Maharaj \cite{sharma2007class} used linear equation of state to establish relativistic compact models consistent with observational data. Physically viable relativistic compact stars models studied by Ngubelanga {\it et al.} \cite{ngubelanga2015relativistic}  for a linear equation of state in isotropic coordinates. For getting a solution of anisotropic distributions with a quadratic equation of state (EOS) deliberated by Sharma and Ratanpal \cite{sharma2013relativistic}, Feroze and Siddiqui \cite{feroze2011charged} and Takisa and Maharaj \cite{takisa2013compact}. Thirukkanesh and Ragel \cite{thirukkanesh2012exact} and Takisa and Maharaj \cite{takisa2013some} have implemented polytropic EOS to generate results for relativistic stars.

Knutsen \cite{knutsen1988some} gave the set-up for general conditions like causality condition, regularity condition and strong and weak energy condition. It has been observed that if the tangential pressure (denoted by $ p_{\perp} $) is higher than the radial pressure (represented by $ p_{r} $), the system becomes more stable. The effect of anisotropy has been derived by Ivanov \cite{ivanov2002static}. Once Einstein's field equations are solved using the energy-momentum tensor, using boundary condition with $ r=R $, one can obtain the constant of integration and eventually determine the mass and radius of the star. In 2007, Sharma and Maharaj \cite{sharma2007class} studied the linear equation of state for relativistic stars choosing $ e^{\lambda}=\frac{1+ar^2}{1+(a-b)r^2} $ in the space-time metric as the coefficient of $ dr^2 $, get the different result for different values of $ a $ and $ b $. Thirukkanesh and Maharaj \cite{thirukkanesh2008charged} have studied charged anisotropic matter with a linear equation of state by specifying a particular form for one of the gravitational potentials and the electric field intensity. Ivanov \cite{ivanov2020linear} has studied generating solutions for linear and Ricatti equation in general relativity.

Felice {\it et al.} \cite{de1999relativistic} and Ray {\it et al.} \cite{ray2003electrically} suggested the models generated which have been used in the description of neutron stars and black hole formation. Several models of charged relativistic matter have been studied by researchers, for example, Komathiraj and Maharaj \cite{komathiraj2007analytical}, Thirukkanesh and Maharaj \cite{thirukkanesh2008charged}. Bare quark stars were considered by Usov {\it et al.} \cite{usov2005structure}, hybrid proto-neutron stars by Nicotra {\it et al.} \cite{nicotra2006hybrid} and strange quark star matter by Dicus {\it et al.} \cite{dicus2008critical} has been used in charged models. In static spherically symmetric spacetimes, the existence of a conformal killing vector is assumed by Esculpi and Aloma \cite{esculpi2010conformal} for the anisotropic relativistic charged matter. Mak and Harko \cite{mak2002exact} have found exact solutions for strange quark matter. Felice {\it et al.} \cite{de1999relativistic} studied a particular solution relating the radial pressure to the energy with a quadratic equation of state. This is a significant advance as the complexity of the model increases significantly due to the radial pressure non-linearity concerning the energy density. But the survey mentioned it mainly suffers from the undesired property of having specificity in the charge density property of the center of the sphere. Malaver \cite{malaver2014strange} represented a relativistic model with the quadratic equation of state with a charged distribution and a gravitational potential $ z(x) $ that depends on an adjustable parameter. Malaver and Daei \cite{malaver2020relativistic} studied the strange quark star model with the quadratic equation of state to integrate the field equation. Sunzu {\it et al.} \cite{sunzu2014charged} describe matter distribution satisfies a linear equation of state consistent with quark matter. Maharaj {\it et al.} \cite{maharaj2014some} derived some simple models for quark stars by considering the charged anisotropic matter with a linear equation of state.

\noindent The objective of this paper is to generate exact solutions to the Einstein field equation, with a linear equation of state that may be derived to analyse charged anisotropic relativistic compact stars. Section 2 expresses the paraboloidal spacetime and the field equations assuming anisotropic matter distributions. In section 3, we derived the model by assuming the linear equation of state $ p=p(\rho) $ by taking ansatz
$ e^{\lambda}=1+\frac{r^2}{R^2} $. In section 4, finding the parameter R, A and $ \alpha $. We have used matching condition for the interior spacetime metric with the Schwarzchild exterior metric across the boundary $ r=a $. In section 5, we have found the bounds on A and $ \alpha $ by using the feasibility conditions at the centre $ r = 0 $ and on the boundary at $ r = a $. We have obtained solutions of EFEs satisfying the linear equation of state on a paraboloidal spacetime compatible with observational data of many compact star candidates like 4U 1820-30, PSR J1903+327, EXO 1785-248, Vela X-1, PSR J1614-2230, Cen X-3. All the physical acceptability conditions have been extensively discussed in section 5. In section 5, we have discussed the physical viability of the model using the graphical method for the compact stars like  4U 1820-30, PSR J1903+327, EXO 1785-248, Vela X-1, PSR J1614-2230, Cen X-3. In section 6, we have concluded by pointing out the main results of our model.

\section{The Spacetime Metric}
\label{sec:2}
\noindent 
A three-paraboloid immersed in a four-dimensional Euclidean space has the cartesian equation
\begin{equation}\label{line element}
	x^2 +y^2 +z^2 = 2wR,
\end{equation} 
where w = constants gives a spheres, while x = constants, y = constants, z = constants respectively, give 3- paraboloids.
on taking the parametrization
\begin{equation*}
	x = r sin\theta cos\phi,
\end{equation*}	
\begin{equation*}
	y = r sin\theta sin\phi,
\end{equation*}	
\begin{equation*}
	z = r cos\theta,
\end{equation*}	
\begin{equation}\label{parametrization}
	w = \frac{r^2}{2R},
\end{equation}		
\\the euclidean metric
\begin{equation}\label{e1}
	d\sigma^2 = dx^2 + dy^2 +dz^2 +dw^2.
\end{equation}
takes the form
\begin{equation}\label{e5}
	ds^2 =\left(1+\frac{r^2}{R^2}\right) dr^2 -r^2(d\theta^2 +sin^2 \theta d\phi^2)
\end{equation}
We shall take the interior spacetime metric for the anisotropic fluid distribution as 
\begin{equation}\label{e2}
	ds^2 = e^\nu dt^2-\left(1+\frac{r^2}{R^2}\right)dr^2 -r^2(d\theta^2 +sin^2 \theta d\phi^2)
\end{equation}

The constant $ \frac{1}{R^2} $  can be identified with the curvature parameter. We take the energy-momentum tensor for an anisotropic-charged imperfect fluid sphere to be of the form
\begin{equation}\label{e4}
	T_{ij}=diag\left(\rho+{E^{2}} , P_{r}-{E^{2}} ,P_{\perp}+{E^{2}} ,P_{\perp}+{E^{2}}\right)
\end{equation}
where $ \rho $ is the matter density, $ P_{r} $ is the radial pressure, $ p_{\perp} $ is the tangential pressure, and E is the electric field intensity. with spacetime metric (\ref{e2}) and energy-momentum technique (\ref{e4}) the Einstein's field equations, takes the form		
\begin{equation}\label{e6}
	8\pi\rho + E^2 = \frac{1-e^{-\lambda}}{r^2}+\frac{e^{-\lambda}\lambda'}{r},
\end{equation} 
\begin{equation}\label{e7}
	8\pi p_{r} - E^2 = \frac{e^{-\lambda}\nu'}{r}+\frac{e^{-\lambda}-1}{r^{2}},
\end{equation}
\begin{equation}\label{e8}
	8\pi p_{\perp}  + E^2=e^{-\lambda} \left(\frac{\nu^{''}}{2} +\frac{\nu^2}{4}-\frac{\nu' \lambda'}{4}+\frac{\nu'-\lambda'}{2r}\right),
\end{equation}
\begin{equation}\label{e21}
	8\pi \Delta = 8\pi P_{r}-8\pi P_{\perp}.
\end{equation} 
where primes denote differentiation with respect to r. The system of equation (\ref{e6}-\ref{e21}) governs the behaviour of the gravitational field for an anisotropic charged fluid distribution.
choosing
\begin{equation}\label{e9}
	E^{2} =\frac{\alpha \frac{r^2}{R^2}}{R^2(1+\frac{r^2}{R^2})^{2}}
\end{equation} 
By substituting the value of $ 	e^{\lambda}  = 1+\frac{r^2}{R^2}  $ and $ E^2 $ in the equation (\ref{e6}), we get
\begin{equation}\label{e13}
	8\pi\rho=\frac{3+(1-\alpha)\frac{r^{2}}{R^{2}}}{R^2(1+\frac{r^2}{R^2})^{2}}
\end{equation}
The expression for density $ \rho(r) $ is finite at the centre of the star.
\section{Linear Equation of state}
\label{sec:3}
\noindent We anticipate that the matter distribution should meet a barotropic equation of state for a physically plausible relativistic star $ p_{r}=p_{r}(\rho) $. Many researchers have presented their idea on the linear equation of state Sharma and Maharaj \cite{sharma2007class}, Thirukkanesh and Maharaj \cite{thirukkanesh2008charged}, Thomas and Pandya \cite{thomas2017anisotropic}. We consider a linear equation of state between the radial pressure $ p_{r} $  and matter density $ \rho $ as
\begin{equation}\label{e14}
	P_{r}=A\rho-B,
\end{equation}
Where A and B are constants. The radius of the star with this pressure distribution is obtained by using the condition 
\begin{equation*}
	P_{r}(r=a) = 0,
\end{equation*}  
gives,
\begin{equation}\label{e15}
	B = A\rho_{a}
\end{equation}	 
We substitute equation (\ref{e15}) in (\ref{e14}) , We get
\begin{equation}\label{e16}
	P_{r}=A\rho-A\rho_{a}=A(\rho-\rho_{a})
\end{equation}
Applying the equation (\ref{e16}) in equation (\ref{e7}), We get
\\					
\begin{equation*}					
	\nu'=re^{\lambda}[A(\rho-\rho{a})-(\frac{e^{-\lambda}-1}{r^{2}})-E^{2}],			
\end{equation*} 
			
\begin{equation}\label{e17}
	\nu'= r\left(1+\frac{r^{2}}{R^{2}}\right)\left(\frac{1}{r^{2}+R^{2}}+ A\left(\frac{-3R^{2}+a^{2}(\alpha-1)}{(a^{2}+R^{2})^{2}}+\frac{3-\frac{r^{2}}{R^{2}}(1-\alpha)}{R^{2}(r^{2}+R^{2})^{2}}-\frac{r^{2}\alpha}{(r^{2}+R^{2})^{2}}\right)\right),
\end{equation}
using equation (\ref{e17}), we get

\begin{equation*}
	e^{\nu}=c\left(r^{2}+R^{2}\right)^{\left(\frac{\alpha+A(2+\alpha)}{2}\right)} 	
\end{equation*}

\begin{equation}\centering
	\times exp\left[{(A+1)(1-\alpha)\frac{r^2}{2R^2}-\frac{A}{2}\left(\left(3+(1-\alpha)\frac{a^2}{R^2}\right)\left(1+\frac{r^2}{2R^2}\right)\left(1+\frac{a^2}{R^2}\right)^{-2}\dfrac{r^2}{R^2}\right)}\right],	
\end{equation}

Where c is a constant of integration. This is a new exact solution for charged linear equation of state. If we put $ \alpha = 0 $	in the expression of $ e^{\nu} $, we get the same expression given by Thomas and Pandya \cite{thomas2017anisotropic}.

\section{Matching Condition}
\label{sec:4}
\noindent The solutions presented in this work might be related to Einstein's field equations. The spacetime metric (\ref{e2}) together with (\ref{parametrization}) should continuously useful with the Reissner-Nordstr{\"o}m exterior spacetime
\begin{equation}
	ds^{2}=\left(1-\frac{2M}{r}+\frac{Q^{2}}{r^{2}}\right)dt^{2}-\left(1-\frac{2M}{r}+\frac{Q^{2}}{r^{2}}\right)^{-1}dr^{2} -r^2(d\theta^2+sin^2\theta d\phi^2)
\end{equation} 
This leads to across the boundary r = a of the star.
We get, 
\begin{flalign}
	M&=\frac{\frac{a^3}{R^2}\left(1+(1+\alpha)\frac{a^2}{R^2}\right)}{2(1+\frac{a^2}{R^2})^2} \\
	c&=\frac{R^2}{(a^2+R^2)^{\frac{(A+1)(2+\alpha)}{2}}} \times F 
	\end{flalign}
where, 
\begin{equation*}
	F = exp\left({\frac{1}{4R^2}\left({AR^2+a^2A\alpha-Aa^2+2a^2\alpha+2A\alpha R^2-2a^2+2R^2\alpha-2R^2}\right)}\right)
\end{equation*}
The expression of matter density, radial pressure and tangential pressure then takes the form
\begin{equation}
	8\pi\rho=\frac{3+(1-\alpha)\frac{r^{2}}{R^{2}}}{R^2(1+\frac{r^2}{R^2})^{2}},
\end{equation}

\begin{equation}\label{e19}
	P_{r}=A\rho-A\rho_{a}=A(\rho-\rho_{a}),
\end{equation}

\begin{equation}\label{e20}
	8\pi P_{\perp}=\frac{-r^2\alpha}{(r^2+R^2)^2}+\frac{A_1+A_2+A_3-A_4+r^2(A_5)^2}{4(1+\frac{r^2}{R^2})}.
\end{equation}
Where
\begin{equation*}
	A_{1}=\frac{-18Ar^2R^2-12AR^4+4R^2a^2A(\alpha-1)+6a^2Ar^2(\alpha-1)}{R^2(a^2+R^2)^2},
\end{equation*}

\begin{equation*}
	A_{2}=\frac{4\alpha R^2+4A(2+\alpha)R^2-4R^2+2(1+A)r^2(\alpha+1)}{R^2(r^2+R^2)^2},
\end{equation*}

\begin{equation*}
	A_{3}=\frac{-6r^2 \alpha -6Ar^2 (2+\alpha)}{(r^2+R^2)^2},
\end{equation*}

\begin{equation*}
	A_{4}=\frac{4(1+A)(\alpha-1)}{R^2},
\end{equation*}

\begin{equation*}
	A_{5}=\frac{A(r^2+R^2)(-3R^2+a^2(\alpha-1))}{R^2(a^2+R^2)^2}-\frac{(1+A)(\alpha-1)}{R^2}+\frac{\alpha}{r^2+R^2}+\frac{A(2+\alpha)}{r^2+R^2}.
\end{equation*}
It connects the variable A, R,$ \alpha $ and c. This shows that there are enough parameters to readily satisfy the persistence of the metric coefficients throughout the boundary of the star $ r = R $. The amount of free parameters readily meets the requirements that develop for a specific model under research. While putting the variables into the expression $ \rho $, $ p_{r} $ and $ p_{\perp} $, we can observe a graphical representation of different stars at the center and boundary of the star in the next section.
\section{Physical Plausibility Condition}
\label{sec:5}
\noindent
Kuchowicz \cite{kuchowicz1972differential}, Buchdahl \cite{buchdahl1979regular}, Knutsen \cite{knutsen1988some} and Murad and Fatema \cite{murad2015some} have given the set of conditions to verify the model is physically justifiable:
(i) Regularity of metric potential
(ii) Radial Pressure and Tangential Pressure at the boundary
(iii) Energy conditions
(iv) Monotone Decrease of Physical Parameters
(v) Pressure Anisotropy
(vi) Surface Redshift
(vii) Stability Conditions
(viii) Mass-Radius Relation
(ix) Stability under three forces acting on the system

\subsection{Regularity of metric potential}
In our metric , at $ r=0 $ , $ e^\lambda = 1 $ and 
\begin{equation*}
	e^{\nu}=c R^{\left(\alpha+A(2+\alpha)\right)} e^{\frac{AR^{2}(-3R^{2}+a^{2}(\alpha-1))-2(A+1)R^{2}(\alpha-1)(a^2+R^2)^2}{4(a^2+R^2)^2}}
\end{equation*}
which are positive constants.
\begin{equation}
	(e^{\lambda})'=\frac{2r}{R^2} , \;\;\;\;  i.e \;\;\; (e^{\lambda})'_{(r=0)} = 0
\end{equation}
\begin{equation}
	(e^\nu)'_{(r=0)} = 0
\end{equation}

Clearly, from the above equation, it shows that metric coefficients are regular at $ r=0 $.

\subsection {Radial Pressure and Tangential Pressure at Boundary}
\begin{figure}[ht]\centering
	\includegraphics[scale = 1.25]{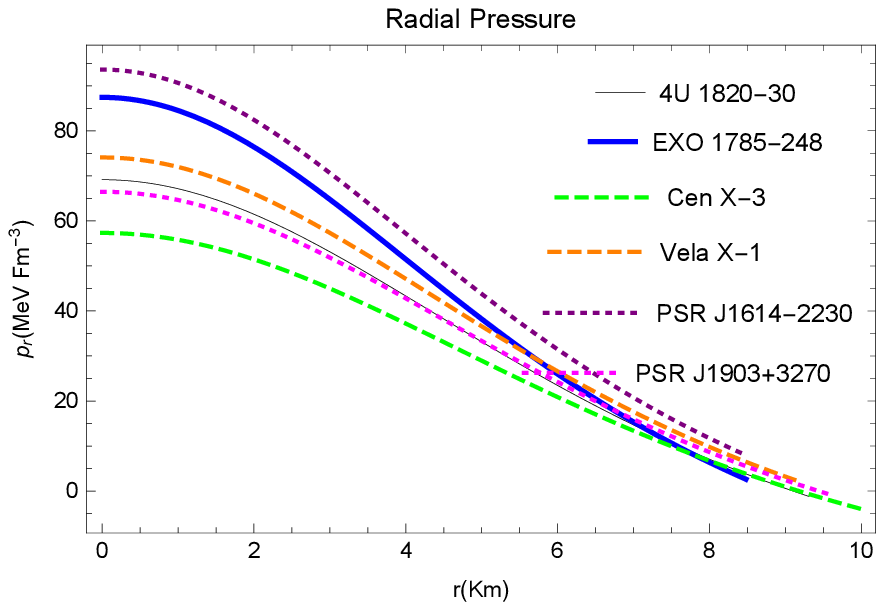}
	\caption{Variation of radial pressures against radial variable $r$.}
	\label{fig:2}
\end{figure}

\begin{figure}[ht]\centering
	\includegraphics[scale = 1.25]{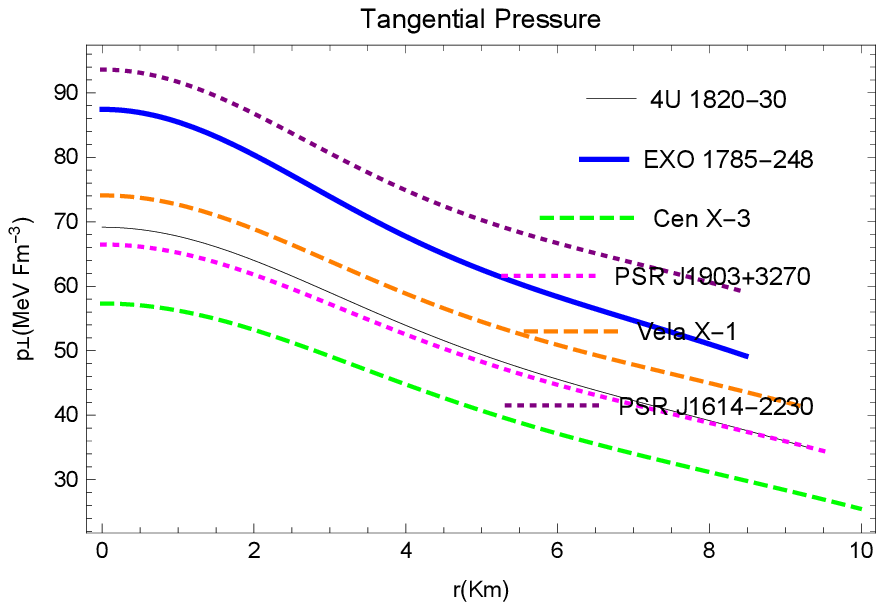}
	\caption{Variation of transverse pressures against radial variable $r$.}
	\label{fig:3}
\end{figure}

The value of radial pressure  $ p_{r} $ should be equal to zero at the surface of the star $ r=a. $ From the equation (\ref{e19}), we can observed that the value of $ p_{r} $  at $ r=a $ is zero. From equations (\ref{e19}) and (\ref{e20}), it can be shown that the conditions $ p_{r}(r = 0)\ge 0 $ , $ p_{\perp}(r = 0) \ge 0 $  and $ p_{\perp}(r = R) \ge 0 $ impose a bound on $ \alpha $, viz.,$  0 \le \alpha \le 0.251789$. It means that these conditions are satisfied in the range $ \alpha $. It can be observed from the graphical method. Fig. \ref{fig:2} and Fig. \ref{fig:3} show that the fact about the condition is satisfied. These conditions are satisfies for the stars  4U 1820-30, PSR J1903+327, EXO 1785-248, Vela X-1, PSR J1614-2230, Cen X-3.

\subsection{Pressure Anisotropy}
\begin{figure}[ht]
\centering
	\includegraphics[scale = 1.25]{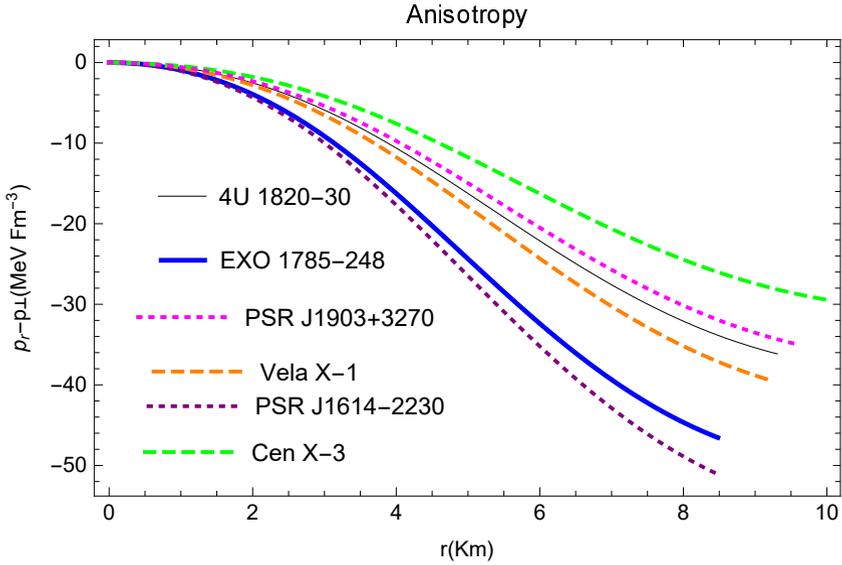}
	\caption{Variation of anisotropies against radial variable $r$.}
	\label{fig:6}
\end{figure}

The difference between tangential and radial pressure should be zero at the centre of a compact star. This condition gives that the pressure components would be equal at a single point. i.e.$ \Delta_{(r=0)}= 0 $ where $ \Delta $ is anisotropy of the star.  Equation (\ref{e21}) is the expression for the anisotropy. Fig. \ref{fig:6} represents the anisotropy for a different star. 

\begin{table}[h]
	\caption{Estimated values of physical parameters based on the observational data for $ \alpha =0.1 $}
	\label{tab:1}
	\begin{tabular}{llllll}
		\hline\noalign{\smallskip}
		\textbf{STAR} & {$ \mathbf{M} $} & {$ \mathbf{a} $} & {$ \mathbf{R} $} & 
		{$ \mathbf{ \rho - p_{r} - 2p_{\perp}}_{(r=0)} $} & {$ \mathbf{\rho-p_{r}-2p_{\perp}}_{(r=a)} $} 
		\\	&  $ \mathbf{(M_\odot)} $ & $ \mathbf{(Km)} $ & \textbf{(Km)} & \textbf{(MeV fm{$\mathbf{^{-3}}$})} & \textbf{(MeV fm{$\mathbf{^{-3}}$})}   \\
		\noalign{\smallskip}\hline\noalign{\smallskip}
		\textbf{4U 1820-30} 	  & 1.58   & 9.1   &  9.31  & 834.813   & 279.409 \\
		\textbf{PSR J1903+327} 	  & 1.66  & 9.438  &  9.54  & 793.228   & 258.666 \\
		\textbf{EXO 1785-248} 	  & 1.3   & 8.849  &  8.48  & 993.993   & 286.58 \\
		\textbf{Vela X-1} 	      & 1.77  & 9.56   &  8.45  & 846.824   & 246.009 \\
		\textbf{PSR J1614-2230}   & 1.97  & 9.69   &  9.19  & 984.427   & 219.801 \\
		\textbf{Cen X-3}          & 1.49  & 9.178  &  9.98  & 809.696   & 275.308 \\
		\noalign{\smallskip}\hline
	\end{tabular} 
\end{table}
In Table \ref{tab:1}, We have calculated the values of strong energy condition for various stars at the boundary $ (r=a) $ and at center $ (r=0) $, which is one of the required conditions to justify the model's feasibility of compact stars. 
\subsection{Monotone Decrease of Physical Parameters}
\begin{figure}[ht]
\centering
	\includegraphics[scale = 1.25]{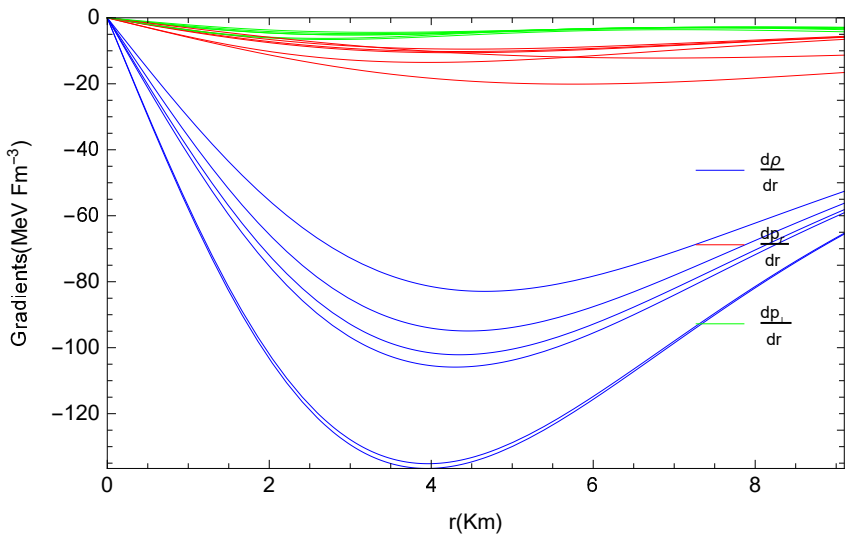}
	\caption{Variation of a Gradients $ \frac{d\rho}{dr} $,$ \frac{dp_{r}}{dr} $ and $ \frac{dp_{\perp}}{dr} $ for different compact stars.}
	\label{fig:11}
\end{figure}

The conditions for monotone decrease of physical parameters are as \;\;$ \frac{d\rho}{dr}\le 0 $ , \;\;$ \frac{dp_{r}}{dr}\le 0 $ \;\; and \;\; $ \frac{dp_{\perp}}{dr}\le 0 $ \;\; for $ 0\le r\le a. $
The gradients of density and radial pressure are given by $ \dfrac{dp_{r}}{dr} = A (8\pi\dfrac{d\rho}{dr}) = A (\frac{2r^3(\alpha-1)-2rR^2(5+\alpha)}{(r^2+R^2)^3}) $ \;\; for $ 0\le r\le a. $ Fig. \ref{fig:11} shows a gradient of density, radial pressure and tangential pressure are decreasing radially outward for the compact stars 4U 1820-30, PSR J1903+327, EXO 1785-248, Vela X-1, PSR J1614-2230, Cen X-3.

\subsection{Energy conditions}
\begin{figure}[ht]
\centering
	\includegraphics[scale = 1.25]{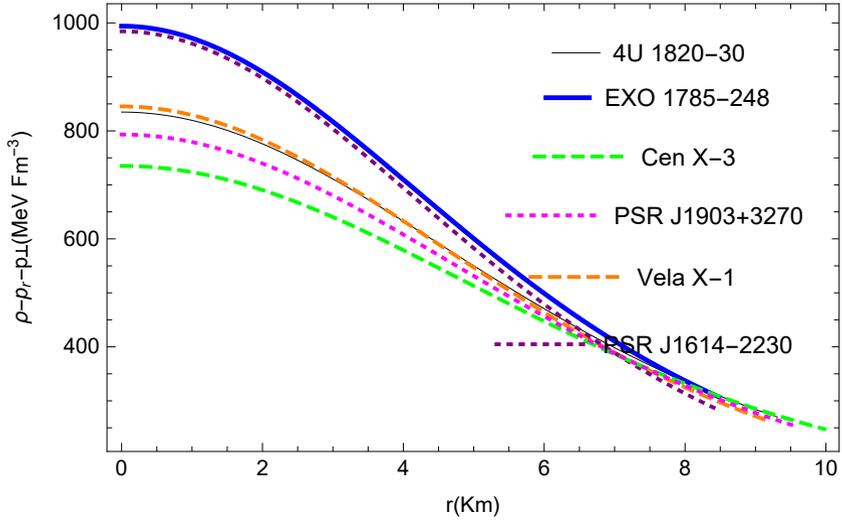}
	\caption{Variation of strong energy condition against radial variable $r$.}
	\label{fig:7}
\end{figure}

(i) $ \rho -p_{r}-2p_{\perp}\ge 0  $ (Strong energy conditions).
\\	The verification of strong energy condition is verified in Table \ref{tab:1} Strong energy conditions
$ \rho - p_{r}-2p_{\perp} \ge 0  $ at the centre $ r = 0 $ and surface of the star $ r = R. $ when we simplify this condition on $ \alpha $, we get the restriction on $ \alpha $ viz., $  0 \le \alpha \le 0.3215 $. The condition is fulfilled within the range of $ \alpha $. Fig. \ref{fig:7} is the evidence for the fulfilment of the condition for different stars  4U 1820-30, PSR J1903+327, EXO 1785-248, Vela X-1, PSR J1614-2230, Cen X-3.
\\(ii) $ \rho \ge p_{r} $ and $ \rho \ge p_{\perp} $ (Weak Energy conditions).
\\	The Weak energy indicates that $\rho - p_{r} \ge 0 $ and $\rho - p_{\perp} \ge 0 $. Since the strong energy condition yields a positive value indicating that $ \rho $ is greater than both $ p_{r} $ and $ p_{\perp}. $ This implies that the difference between $ \rho $ and $ p_{r} $ as well as $ \rho $ and $ p_{\perp} $ is greater than equal to zero.

\subsection{Stability Conditions}
\begin{figure}[ht]\centering
	\includegraphics[scale=1.25]{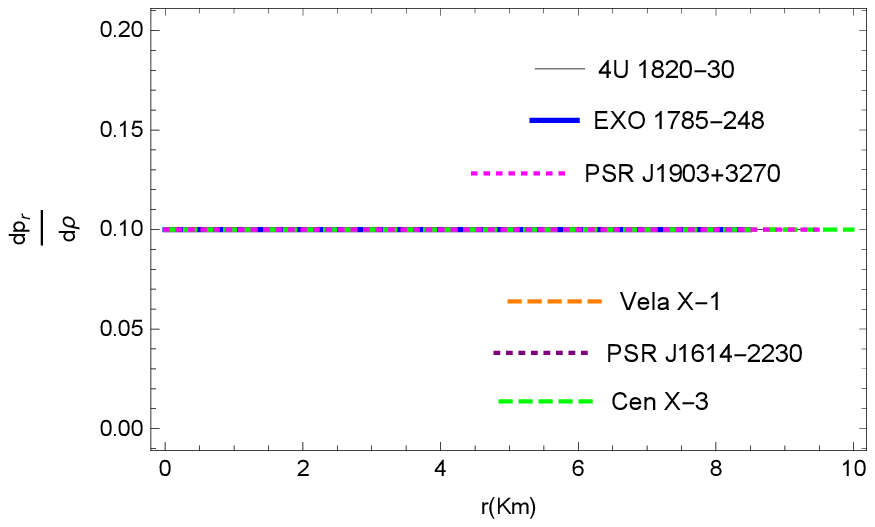}
	\caption{Variation of $ \frac{dp_r}{d\rho} $ against radial variable $r$.}
	\label{fig:4}
\end{figure}

\begin{figure}[ht]\centering
	\includegraphics[scale = 1.25]{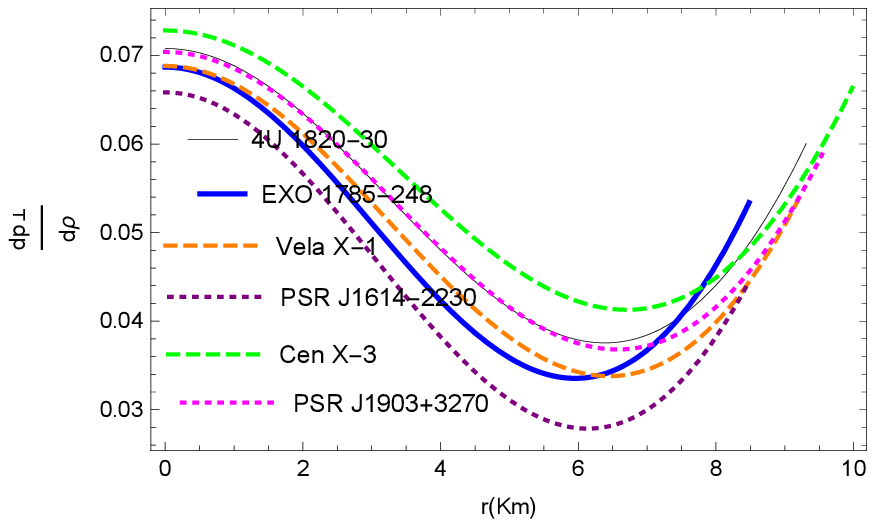}
	\caption{Variation of $ \frac{dp_\perp}{d\rho} $ against radial variable $r$.}
	\label{fig:5}
\end{figure}

\begin{figure}[ht]\centering
	\includegraphics[scale = 1.25]{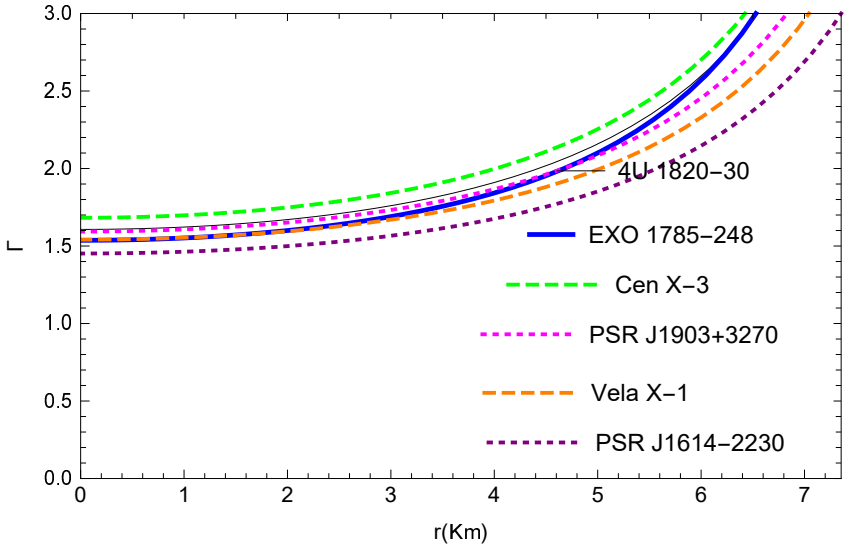}
	\caption{Variation of adiabatic Index against radial variable $r$.}
	\label{fig:8}
\end{figure}

(i) Causality condition:
$ 0\le \frac{dp_{r}}{d\rho}\le 1, \;\; 0\le \frac{dp_{\perp}}{d\rho}\le 1  $ \;\;  for $ 0\le r \le a. $
\\ The values for the radial speed of sound waves  $ \frac{dp_{r}}{d\rho} $ denoted as $ \nu^{2}_{r} $ and transverse speed of sound waves $ \frac{dp_{\perp}}{d\rho} $ denoted as $ \nu^{2}_{t} $ between $ r = 0 $ and $ r = a $ for different stars have been calculated in Table \ref{tab:3}. These velocities are in the range of 0 and 1.
The bounds on $ \alpha $ for $ 0\le \frac{dp_{r}}{d\rho}\le 1, \;\; 0\le \frac{dp_{\perp}}{d\rho}\le 1  $ are $ 0 \le \alpha \le 2.28165 \;\; and \;\; 0 \le \alpha \le 1.11724 $. Fig. \ref{fig:4}  and Fig. \ref{fig:5} show that these conditions are satisfied throughout the distribution.
\\(ii) Relativistic adiabatic index:
$ \Gamma =(\frac{\rho +p_{r}}{p_{r}}) \frac{dp_{r}}{d\rho} $
\\ The adiabatic index stated must be greater than 1.333... in the prescribed range $ 0 \le r \le a. $ $\Gamma \ge \frac{4}{3}  $ at $ r = 0  $  imposes a restriction on $ \alpha $ given by $ 0 \le \alpha \le 1.8686 $. Table \ref{tab:3} shows these respective values for the different compact stars. Fig. \ref{fig:8}  shows that these conditions are satisfied throughout the distribution.

\subsection{Redshift}
\begin{figure}[ht]
\centering
	\includegraphics[scale = 1.25]{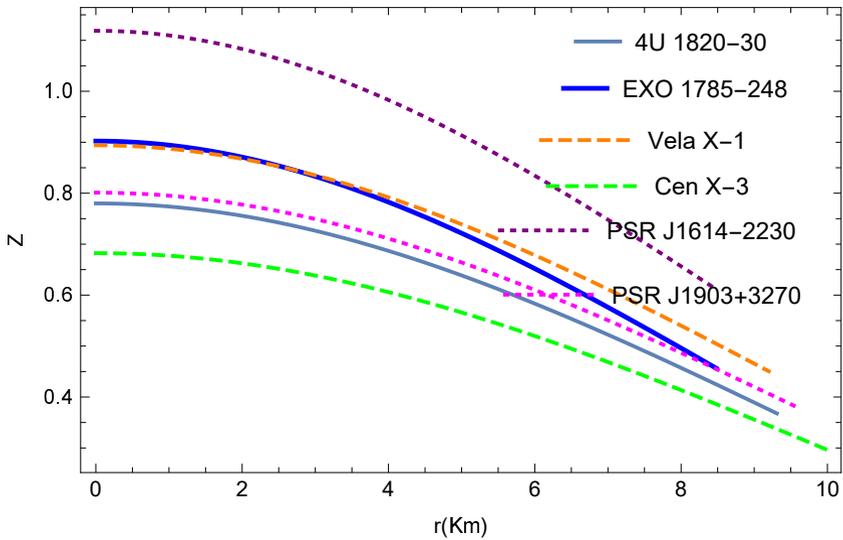}
	\caption{Variation of surface redshift against radial variable $r$.}
	\label{fig:9}
\end{figure}

The redshift $ z = \sqrt{1/e^{\nu}} -1 $ must be a decreasing function of r and finite for $ 0\le z \le 5 $.
In Table \ref{tab:2}, we have described all the values for different compact stars (related to the stability of a relativistic anisotropic stellar configuration). The value of the redshift remains less than 5.  For a relativistic star, it is expected that the redshift must decrease towards the boundary and be finite throughout the distribution. Fig. \ref{fig:9} shows that gravitational redshift decreases throughout the star under consideration for  4U 1820-30, PSR J1903+327, EXO 1785-248, Vela X-1, PSR J1614-2230, Cen X-3.

\subsection{Mass-Radius Relation:}
\begin{figure}[ht]
\centering
	\includegraphics[scale = 1.25]{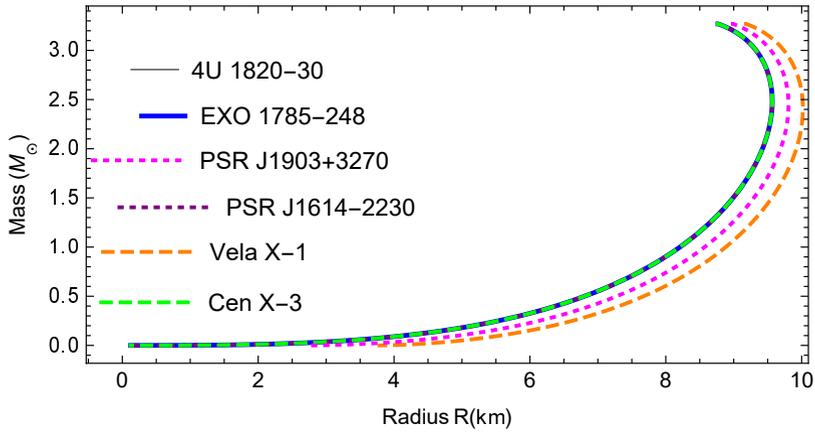}
	\caption{Variation of a mass M with respect to a radius R for various stars.}
	\label{fig:10}
\end{figure}

According to Buchdahl \cite{buchdahl1979regular}, the mass radius relation must satisfy the inequality, $ \frac{M}{a}\le  \frac{4}{9} $
The values are calculated in Table \ref{tab:2} to verify this inequality.
We can verify this Mass-Radius Relation with graphical Method $ \frac{M}{a} = 0.256 < \frac{4}{9}. $  Fig. \ref{fig:10} shows the Mass-Radius Relation for the star  4U 1820-30, PSR J1903+327, EXO 1785-248, Vela X-1, PSR J1614-2230 and Cen X-3.

\subsection{Stability under three forces acting on the system}
\begin{figure}[ht]
\centering
	\includegraphics[scale = 1.25]{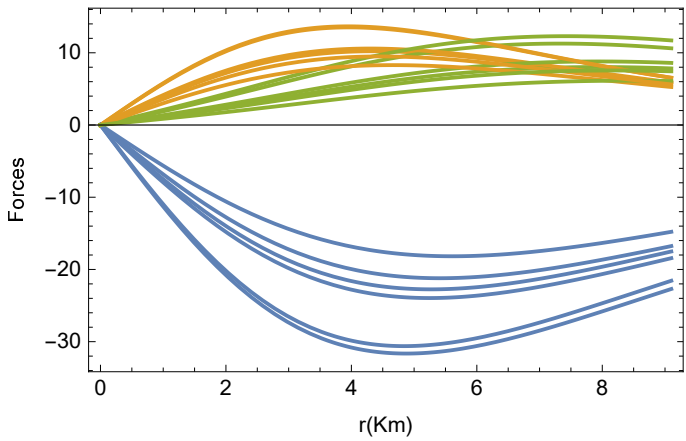}
	\caption{Variation of three forces like Gravitational Force(Blue), Hydrostatic Force(Orange) and Anisotropic Force(Green) for different compact star.}
	\label{fig:12}
\end{figure}

We want to examine the stability of our present model under three different forces
viz., gravitational force, hydrostatics force and anisotropic force, which the following equation can describe
\begin{equation}\label{TOV}
	-\frac{M_{G}(r)(\rho+p_{r})}{2}e^{(\nu-\lambda)/2}-\dfrac{dp_{r}}{dr}-\frac{2}{r}(p_{\perp}-p_{r})=0
\end{equation}
proposed by Tolman-Oppenheimer-Volkov and named as TOV equation.
The quantity $M_{G}(r)  $  represents the gravitational mass within the radius r, which
can be derived from the Tolman-Whittaker formula and Einstein’s field equations
and is defined by
\begin{equation}\label{mgr}
	M_{G}(r)= \frac{\nu'}{2}re^{(\nu-\lambda)/2}
\end{equation}
Adding the value of (\ref{mgr}) into equation (\ref{TOV}), we get,
\begin{equation}\label{TOV1}
	-\frac{\nu'}{2}re^{(\nu-\lambda)/2}(\rho+p_{r})e^{(\nu-\lambda)/2}-\dfrac{dp_{r}}{dr}-\frac{2}{r}(p_{\perp}-p_{r})=0
\end{equation}
The above expression may also be written as
\begin{align}
		F_{g}+F_{h}+F_{a}=0 \label{TOV2} \\
		F_{g}=	-\frac{\nu'}{2}r(\rho+p_{r}) \label{TOV3} \\
		F_{h}=	-\dfrac{dp_{r}}{dr} \label{TOV4} \\
		F_{a}=	-\frac{2}{r}(p_{\perp}-p_{r}) \label{TOV5}
\end{align} 
5
the three different forces act on the system. The figure shows that gravitational 5force is negative and dominating in nature which is counterbalanced by the combined effect of hydrostatics and anisotropic forces to keep the system in equilibrium. Fig. \ref{fig:12} shows these three forces in graphical method for the star  4U 1820-30, PSR J1903+327, EXO 1785-248, Vela X-1, PSR J1614-2230, Cen X-3.
\begin{table}[h]
	\caption{Estimated values of physical parameters based on the observational data for $ \alpha =0.1 $}
	\label{tab:2}
	\footnotesize
	\begin{tabular}{lllllll}
		\hline\noalign{\smallskip}
		\textbf{STAR}   &  {$ \mathbf{M} $} & {$ \mathbf{a} $} & {$ \mathbf{Z_{r=0}} $} & {$ \mathbf{ Z_{(r=a)}} $} &  {$ \mathbf{\Gamma_{(r=0)}}$}  & {$ \mathbf{u (=\frac{M}{a})} $}  \\
		&  $ \mathbf{(M_\odot)} $ & $ \mathbf{(Km)} $ & \textbf{(Redshift)} & \textbf{(Redshift)} & \textbf{(Adiabatic }  & \textbf{(Buchdahl } \\
		
		&  $ \mathbf{} $ & $ \mathbf{} $ & \textbf{} & \textbf{} & \textbf{ Index)}  & \textbf{ Ratio)} \\
		\noalign{\smallskip}\hline\noalign{\smallskip}
		\textbf{4U 1820-30} 	  & 1.58   & 9.1   &  0.800561  & 0.398355   & 1.60715 & 0.173\\
		\textbf{PSR J1903+327} 	  & 1.66  & 9.438  &  0.822543  & 0.406674    & 1.59342 & 0.175 \\
		\textbf{EXO 1785-248} 	  & 1.3  & 8.849   &  0.902282  & 0.425903   & 1.53685 & 0.146\\
		\textbf{Vela X-1} 	      & 1.77  & 9.56   &  0.921587  & 0.442963   & 1.53998 & 0.185 \\
		\textbf{PSR J1614-2230}   & 1.97  & 9.69   & 1.1548  & 0.521521   & 1.45172 & 0.203\\
		\textbf{Cen X-3}          & 1.49  & 9.178  &  0.787317   & 0.393294    & 1.61581 & 0.162 \\
		\noalign{\smallskip}\hline
	\end{tabular} 
\end{table}

\begin{sidewaystable}
	\sidewaystablefn%
	\begin{center}
		\begin{minipage}{\textwidth}
		\caption{Estimated values of physical parameters based on the observational data for $ \alpha =0.1 $}
		\label{tab:3}
		\begin{tabular}{lllllllll}
			\hline\noalign{\smallskip}
			\textbf{STAR} &  {$ \mathbf{M} $} & {$ \mathbf{a} $} &  {$ \mathbf{\left(\frac{dp_{r}}{d\rho}\right)}_{r=0}  $} & {$ \mathbf{\left(\frac{dp_{\perp}}{d\rho}\right)_{r=0}} $} & {$ \mathbf{(\nu^{2}_{t}-\nu^{2}_{r})_{r=0}} $} & {$ \mathbf{\left(\frac{dp_{r}}{d\rho}\right)_{r=a}} $} & {$ \mathbf{\left(\frac{dp_{\perp}}{d\rho}\right)_{r=a}} $} & {$ \mathbf{(\nu^{2}_{t}-\nu^{2}_{r})_{r=a}} $}\\
			&  $ \mathbf{(M_\odot)} $ & $ \mathbf{(Km)} $ & \textbf{} & \textbf{} & \textbf{}   \\
			\noalign{\smallskip}\hline\noalign{\smallskip}
			\textbf{4U 1820-30} 	  & 1.58   & 9.1   &  0.1  & 0.0707929  & -0.0292071 & 0.1 & 0.0567916 & -0.043 \\
			\textbf{PSR J1903+327} 	  & 1.66  & 9.438  &  0.1  & 0.0703981   & -0.0296019 & 0.1 & 0.0572536 & -0.042 \\
			\textbf{EXO 1785-248} 	  & 1.3  & 8.48   &  0.1  & 0.0686885   & -0.0313115 & 0.1 & 0.0599306 & -0.043 \\
			\textbf{Vela X-1} 	      & 1.77  & 9.56   &  0.1  & 0.0658326   & -0.0312131 & 0.1 & 0.442963  & -0.040 \\
			\textbf{PSR J1614-2230}   & 1.97  & 9.69   &  0.1  & 0.0687869   & -0.0341674 & 0.1 & 0.0675101 &-0.032 \\
			\textbf{Cen X-3}          & 1.49  & 9.178  &  0.1  & 0.071038   & -0.028962 & 0.1 & 0.0565314 & -0.048 \\
			\noalign{\smallskip}\hline
		\end{tabular} 	
		\end{minipage}
	\end{center}
\end{sidewaystable}

\section{Discussion}
\noindent We have studied the compatibility of the model developed using Linear Equation of state in the background of paraboloidal spacetime for compact stars like 4U 1820-30, PSR J1903+327, EXO 1785-248, Vela X-1, PSR J1614-2230, Cen X-3. Our model satisfies the elementary physical requirements for representing a superdense compact star through the graphical method. It is found that the model can accommodate the mass and radius of the compact star candidates given by Gangopadhyay {\it et al.} \cite{gangopadhyay2013strange}. It is found that stars whose compactness is more accommodate more density, pressure and anisotropy. The redshift increase with compactness while the value of the adiabatic Index decreases with compactness showing that the stability decreases with an increase in compactness. Pertinent feature of the model is that the exact solution obtained is simple, which is only found in some solutions. We have displayed the physical analysis only for a few compact star models here, but it can be applied to a larger class of known pulsars. The model possesses a definite background spacetime geometry, namely paraboloidal geometry, and the expression involved in the solution is exponential.
\noindent It has been concluded that a large number of pulsars with known masses and radii can be accommodated in the present model, satisfying the linear equation of state.
\pagebreak

\section*{Acknowledgement}
BSR would like to thank IUCAA, Pune, for the facilities and hospitality provided to him where part of the work was carried out.
\\
\bibliography{sn-article_linearapss.bib}

\begin{figure}[ht]\centering
	\includegraphics[scale = 1.25]{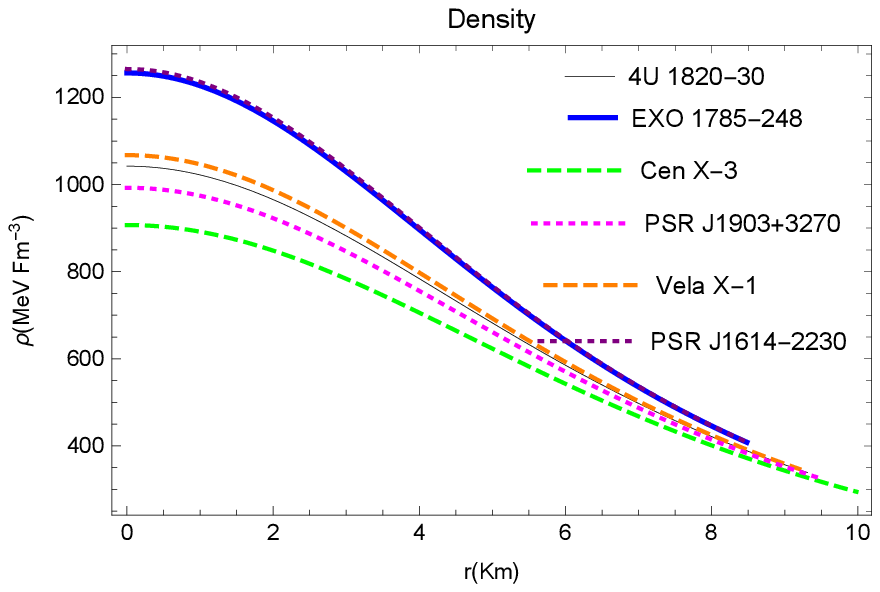} 
	\caption{Variation of density against radial variable $r$.}
	\label{fig:1}
\end{figure}


\end{document}